\documentclass[letter, longauth, traditabstract]{aa}
\usepackage{natbib}
\bibpunct{(}{)}{;}{a}{}{,} 
\usepackage{xspace}
\usepackage{amssymb}
\usepackage{amsmath}
\usepackage{graphicx}
\usepackage{comment}
\usepackage{txfonts}

\def\aap{A\& A}

\def\apjl{ApJL}
\def\apjs{ApJS}

\def\meth{CH$_3$OH}
\def\lsol{L$_\odot$}

\def\mathu{_{\textrm{u}}}

\newcommand{\changes}[1]{#1}
\newcommand{\changestwo}[1]{#1}
\begin{document}

   \author{
Kama M. \inst{\ref{uva}} \and
Dominik C. \inst{\ref{uva},\ref{nijmegen}} \and
Maret S. \inst{\ref{laog}} \and
van der Tak F. \inst{\ref{sron},\ref{kaypten}} \and
Caux E. \inst{\ref{cesr},\ref{cesr2}} \and
Ceccarelli C. \inst{\ref{laog},\ref{bordeaux},\ref{bordeaux2}} \and
Fuente A. \inst{\ref{oan}} \and
Crimier N. \inst{\ref{laog},\ref{cab}} \and
Lord S. \inst{\ref{ipac}} \and
Bacmann A.\inst{\ref{laog},\ref{bordeaux},\ref{bordeaux2}} \and
Baudry A. \inst{\ref{bordeaux},\ref{bordeaux2}} \and
Bell T. \inst{\ref{caltech}} \and
Benedettini M. \inst{\ref{ifsi}} \and
Bergin E.A. \inst{\ref{annarbor}} \and
Blake G.A. \inst{\ref{caltech}} \and
Boogert A. \inst{\ref{ipac}} \and
Bottinelli S. \inst{\ref{cesr},\ref{cesr2}} \and
Cabrit S. \inst{\ref{lerma}} \and
Caselli P. \inst{\ref{leeds}} \and
Castets A. \inst{\ref{laog},\ref{bordeaux},\ref{bordeaux2}} \and
Cernicharo J. \inst{\ref{cab}} \and
Codella C. \inst{\ref{arcetri}} \and
Comito C. \inst{\ref{bonn}} \and
Coutens A. \inst{\ref{cesr},\ref{cesr2}} \and
Demyk K. \inst{\ref{cesr},\ref{cesr2}} \and
Encrenaz P. \inst{\ref{lerma}} \and
Falgarone E. \inst{\ref{lerma}} \and
Gerin M. \inst{\ref{lerma}} \and
Goldsmith P.F. \inst{\ref{jpl}} \and
Helmich F. \inst{\ref{sron}} \and
Hennebelle P. \inst{\ref{lerma}} \and
Henning T. \inst{\ref{heidelberg}} \and
Herbst E. \inst{\ref{ohio}} \and
Hily-Blant P. \inst{\ref{laog}} \and
Jacq T. \inst{\ref{bordeaux},\ref{bordeaux2}} \and
Kahane C. \inst{\ref{laog}} \and
Klotz A. \inst{\ref{cesr},\ref{cesr2}} \and
Langer W. \inst{\ref{jpl}} \and
Lefloch B. \inst{\ref{laog}} \and
Lis D. \inst{\ref{caltech}} \and
Lorenzani A. \inst{\ref{arcetri}} \and
Melnick G. \inst{\ref{cfa}} \and
Nisini B. \inst{\ref{oar}} \and
Pacheco S. \inst{\ref{laog}} \and
Pagani L. \inst{\ref{lerma}} \and
Parise B. \inst{\ref{bonn}} \and
Pearson J. \inst{\ref{jpl}} \and
Phillips T. \inst{\ref{caltech}} \and
Salez M. \inst{\ref{lerma}} \and
Saraceno P. \inst{\ref{ifsi}} \and
Schilke P. \inst{\ref{bonn},\ref{koln}} \and
Schuster K. \inst{\ref{iram}} \and
Tielens X. \inst{\ref{leiden}} \and
van der Wiel M.H.D. \inst{\ref{sron},\ref{kaypten}} \and
Vastel C. \inst{\ref{cesr},\ref{cesr2}} \and
Viti S. \inst{23} \and
Wakelam V. \inst{\ref{bordeaux},\ref{bordeaux2}} \and
Walters A. \inst{\ref{cesr},\ref{cesr2}} \and
Wyrowski F. \inst{\ref{bonn}} \and
Yorke H. \inst{\ref{jpl}}
Cais,	 P. \inst{\ref{bordeaux}} \and
G\"{u}sten, R. \inst{\ref{bonn}} \and
Philipp, S. \inst{\ref{bonn}} \and
Klein, T. \inst{\ref{bonn}} \and
Helmich, F. \inst{\ref{sron}}
          }

   \institute{
Astronomical Institute 'Anton Pannekoek', University of Amsterdam, Amsterdam, The Netherlands
\label{uva}
\and Department of Astrophysics/IMAPP, Radboud University Nijmegen,  Nijmegen, The Netherlands
\label{nijmegen}
\and Laboratoire d'Astrophysique de Grenoble, UMR 5571-CNRS, Universit\'e Joseph Fourier, Grenoble, France
\label{laog}
\and SRON Netherlands Institute for Space Research, Groningen, The Netherlands
\label{sron}
\and Kapteyn Astronomical Institute, University of Groningen, The Netherlands
\label{kaypten}
\and Centre d'\'{E}tude Spatiale des Rayonnements, Universit\'e Paul Sabatier, Toulouse, France
\label{cesr}
\and CNRS/INSU, UMR 5187, Toulouse, France
\label{cesr2}
\and Universit\'{e} de Bordeaux, Laboratoire d'Astrophysique de Bordeaux, Floirac, France
\label{bordeaux}
\and CNRS/INSU, UMR 5804, Floirac cedex, France
\label{bordeaux2}
\and IGN Observatorio Astron\'{o}mico Nacional, Alcal\'{a} de Henares, Spain\label{oan}
\and Centro de Astrobiolog\`{\i}a, CSIC-INTA, Madrid, Spain
\label{cab}
\and Infared Processing and Analysis Center,  Caltech, Pasadena, USA
\label{ipac}
\and California Institute of Technology, Pasadena, USA
\label{caltech}
\and INAF - Istituto di Fisica dello Spazio Interplanetario, Roma, Italy
\label{ifsi}
\and School of Physics and Astronomy, University of Leeds, Leeds UK
\label{leeds}
\and Laboratoire d'\'{E}tudes du Rayonnement et de la Mati\`ere en Astrophysique, UMR 8112  CNRS/INSU, OP, ENS, UPMC, UCP, Paris, France
\label{lerma}
\and Max-Planck-Institut f\"{u}r Radioastronomie, Bonn, Germany
\label{bonn}
\and INAF Osservatorio Astrofisico di Arcetri, Florence Italy
\label{arcetri}
\and Jet Propulsion Laboratory,  Caltech, Pasadena, CA 91109, USA
\label{jpl}
\and Ohio State University, Columbus, OH, USA
\label{ohio}
\and Harvard-Smithsonian Center for Astrophysics, Cambridge MA, USA
\label{cfa}
\and Johns Hopkins University, Baltimore MD,  USA
\label{hopkins}
\and Physikalisches Institut, Universit\"{a}t zu K\"{o}ln, K\"{o}ln, Germany
\label{koln}
\and Institut de RadioAstronomie Millim\'etrique, Grenoble, France
\label{iram}
\and Leiden Observatory, Leiden University, Leiden, The Netherlands
\label{leiden}
\and Department of Physics and Astronomy, University College London, London, UK
\label{ucl}
\and INAF - Osservatorio Astronomico di Roma, Monte Porzio Catone, Italy
\label{oar}
\and Department of Astronomy, University of Michigan, Ann Arbor, USA
\label{annarbor}
\and Max-Planck-Institut f\"ur Astronomie, Heidelberg, Germany
\label{heidelberg}
          }

\title{The methanol lines and hot core of OMC2-FIR4, an intermediate-mass protostar, with {\emph{Herschel}}-HIFI\thanks{\emph{Herschel} is an ESA
       space observatory with science instruments provided by
       European-led principal Investigator consortia and with
       important participation from NASA}}

\titlerunning{Methanol in OMC2-FIR4 as seen by HIFI}
\authorrunning{Kama, CHESS}

\date{DRAFT, \today} 

\abstract{In contrast with numerous studies on the physical and chemical structure of low- and high-mass protostars, much less is known about their intermediate-mass counterparts, a class of objects that could help to elucidate the mechanisms of star formation on both ends of the mass range. We present the first results from a rich HIFI spectral dataset on an intermediate-mass protostar, OMC2-FIR4, obtained in the CHESS (Chemical \textit{HErschel} SurveyS of star forming regions) key programme. The more than 100 methanol lines detected between 554 and 961 GHz cover a range in upper level energy of 40 to 540 K. \changes{Our physical interpretation focusses on the hot core, but likely the cold envelope and shocked regions also play a role in reality, because an analysis of the line profiles suggests the presence of multiple emission components. An upper limit of $10^{-6}$ is placed on the methanol abundance in the hot core, using a population diagram, large-scale source model and other considerations. This value is consistent with abundances previously seen in low-mass hot cores. Furthermore, the highest energy lines at the highest frequencies display asymmetric profiles, which may} arise from infall around the hot core.}


\maketitle

\section{Introduction}\label{sec:introduction}

\changes{Intermediate-mass, and therefore intermediate-luminosity protostars offer insights into the physical and chemical differences between the formation of low- and high-mass stars, but questions about their chemistry and dominant gas-heating mechanisms remain only partly answered.}

\changes{Deep in the interior of a protostellar core, energy is released by a forming protostar. This energy heats the surrounding gas by dust-mediated and UV photon heating, as well as through shocks caused by protostellar outflows. The respective roles of these mechanisms as a function of protostar luminosity and the effects of heating on protostar evolution, are under intense study \citep[e.g.][]{Spaansetal1995, Dotyetal2006, Brudereretal2009}, with the role of shocks and UV photons in low- and intermediate-luminosity sources emphasized by recent \textit{Herschel} Space Observatory results \citep{Fichetal2010, vanKempenetal2010}. Sub-millimeter molecular line emission is a versatile probe of the physical and chemical conditions in these heated regions, the \textit{hot cores}, revealing the initial conditions for forming stars and planetary systems. Hot cores are compact (sizes $<0.1$pc), warm ($\rm T>100$K), and show evidence of complex chemistry \citep{Kurtzetal2000, ceccarelli:ppv}. The two main paths to this complexity are gas-phase and grain-surface reactions \citep[e.g.][]{garrod}. 
During the gradual warm-up of grains in a hot core, species formed on them in earlier evolutionary phases react and the products are later released into the gas phase. Observable chemical differences include a methanol--formaldehyde abundance ratio, which increases with decreasing protostar luminosity \citep{cazaux,bottinelli04,bottinelli07}.}

We present \textit{Herschel}-HIFI sub-millimeter observations of methanol line emisson toward the intermediate-mass protostar \object{OMC2-FIR4}\footnote{\changes{SIMBAD entry: [MWZ90] OMC-2 FIR 4}}, attributing part of the emission to a hot core. With a luminosity of 1000\ \lsol\ \changes{\citep[hereafter Crim09]{Crimieretal2009}} and a distance of only 440 pc \changes{\citep{Hirotaetal2007}}, the protostar OMC2-FIR4 is an excellent laboratory to study hot core chemistry in the intermediate mass regime. \changes{A structure model, constrained by 7.5$''$ to 14.8$''$ resolution dust-continuum maps and the broadband spectral energy distribution, was made for OMC2-FIR4 by Crim09.} \changestwo{The uncertainties in the properties of the central component are large, but the model suggests a hot core radius of $\rm R_{Crim} = 440$} AU and densities above $\rm\sim 5\cdot 10^{6}\ cm^{-3}$. At $\sim$1$''$ resolution, FIR4 is seen to consist of several clumps, which may be forming a cluster of protostars \citep[hereafter Shim08]{Shimajirietal2008}. Some of this activity may be triggered by an outflow from the nearby source \object{OMC2-FIR3}.

Methanol is a powerful diagnostic of the physical and chemical conditions in protostellar sources \citep{vanderTaketal2000, Leurinietal2004, Leurinietal2007, Maretetal2005, Wangetal2010}\changes{, and is used to that end here}. A careful analysis of methanol is also important for recovering the other species which \meth\ lines often blend with.

	\begin{figure}[!t]
	\begin{center}
		\includegraphics[clip=,width=1.0\linewidth]{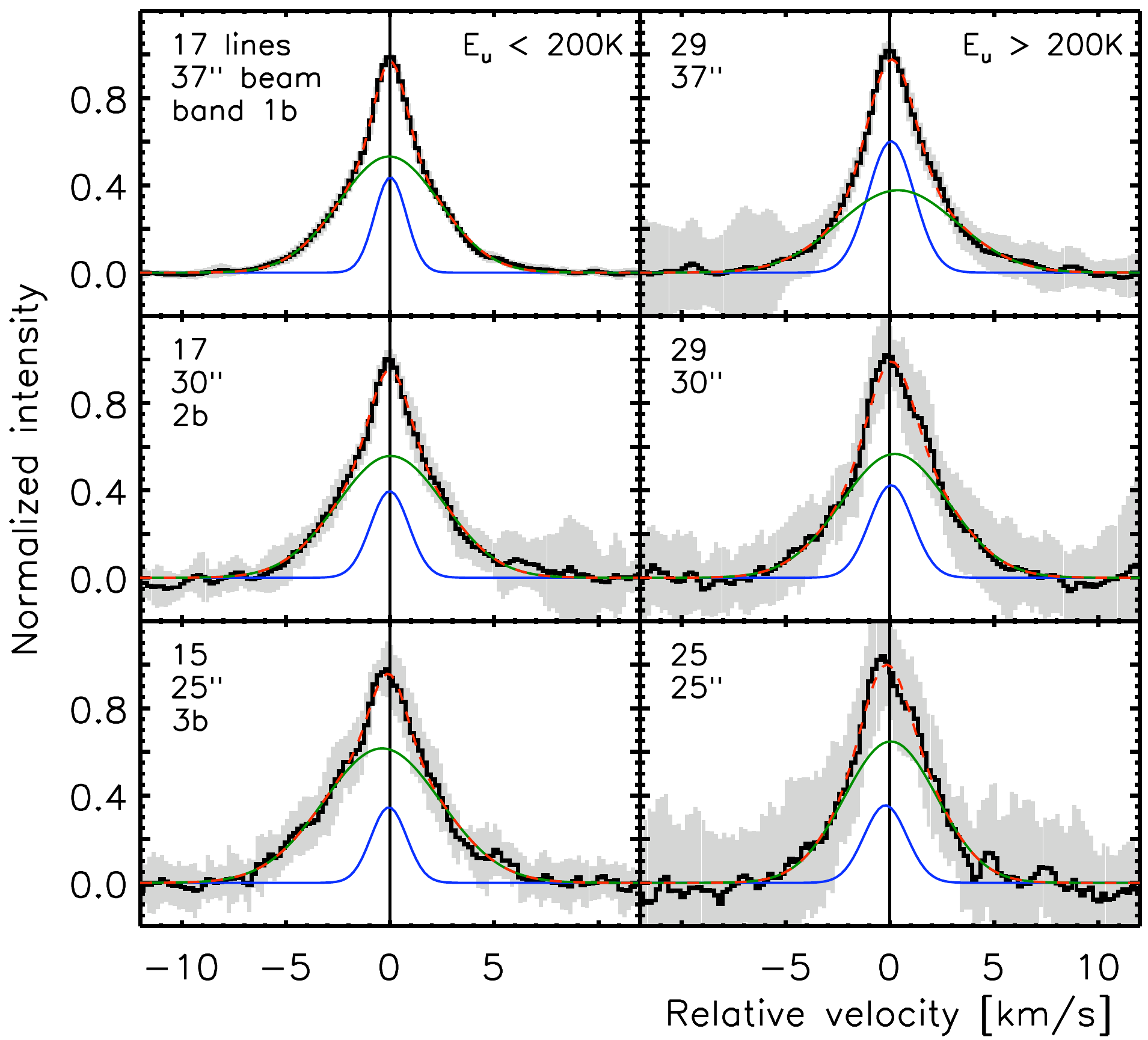}
		\caption{\changes{Averaged \meth\ line profiles in the $\rm E_{u} \leq 200K$ (left) and $\rm E_{u} \geq 200K$ (right) ranges, for bands 1b (top), 2b (middle) and 3b (bottom). The gray area marks the biased weighted standard deviation of the sample of normalized profiles. The two fitted Gaussians (blue, green) and their sum (dashed red line) are given.}}
		\label{fig:lineprofiles}
	\end{center}
	\end{figure}

\section{Observations}\label{sec:observations}

The source OMC2-FIR4 was observed with the HIFI spectrograph on the ESA \textit{Herschel} Space Observatory \citep{Pilbrattetal2010}, \changes{using only the} wide band spectrometer (WBS) \citep{deGraauwetal2010}, covering bands 1b (554.5 to 636.5 GHz), 2b (714.1 to 801 GHz), and 3b (858.0 to 961.0 GHz). The WBS native resolution is \changes{$1.1$MHz}. The data were taken in dual beam switch (DBS) spectral scan mode with a redundancy of four. More observations, expected later in 2010, will cover most of HIFI's frequency range. \changes{Median baseline subtraction and sideband deconvolution were performed with the HIPE 2 software \citep{ott2010}. A frequency-dependent main beam efficiency within three percent of 0.70\changestwo{ and a calibration uncertainty of 15\%} were used in calculating line fluxes. For a detailed review of HIFI's in-orbit performance, see \citet{Roelfsemaetal2010}.}

The data quality in 1b is superb, whereas 2b and 3b are significantly influenced by \changes{spurious features, ``spurs'', introduced by the local oscillator electronics}, which act to increase noise in the single-sideband spectra and can damage spectral line profiles. Combining the H and V polarizations, the RMS outside the bright lines is \changes{T$_{a}^{\star}=10$ mK in 1b, 21 mK in 2b, and 26 mK in 3b.}

\section{Results}\label{sec:results}

Preliminary data processing reveals 91 lines from 17 species in band 1b, and similar numbers in 2b and 3b, establishing OMC2-FIR4 as a relatively line-rich protostar (see also the review by Ceccarelli et al. in this volume). Methanol lines are advantageous to study in OMC2-FIR4 because they are abundant, cover a large range of excitation conditions, and are straightforward to measure because the spectrum is less crowded than in higher luminosity \changes{hot sources, in which more levels of more species can be significantly excited}. Out of the hundreds \changes{of \meth\ transitions in each band, we detect 46 lines} in band 1b, 46 in 2b, and 40 in 3b, with a range of $\rm 40\ K \leq E_{u} \leq 540$ K in excitation energy. Identification made use of the JPL database\footnote{The catalogue is available on-line at http://spec.jpl.nasa.gov.}, with methanol data by \citet{Xuetal2008}. \changes{This paper discusses lines detected with $> 4\sigma$ confidence and not blended or damaged by spurs.}

\subsection{Line profiles}

 \changes{Average line profiles were calculated in two energy ranges, $\rm E_{u} \leq 200$ K and $\rm E_{u} \geq 200$ K}, for each observed band. \changes{These six average profiles are presented in Fig.~\ref{fig:lineprofiles}, with the number of lines and relevant beam sizes given. Before averaging, each line profile was resampled to an 0.2 km/s resolution grid with origin at the rest frequency, fitted by a single Gaussian and normalized by the intensity at the fit center. Distinguishing between the core ($\rm |v| < 2$ km/s) and wings ($\rm |v| > 2$ km/s) of a profile, we note that the cores of the average line profiles seem to show a broadening trend with decreasing beam size and increasing E$\mathu$. Also, the wings are weaker in the higher energy regime than in the lower.}

\begin{table}[!t]
\begin{center}
\begin{tabular}{ r  l  r@{ $\pm$ }l }
\hline \hline
E$\mathu$ range	&	Fit	&	\multicolumn{2}{c}{FWHM [km/s] (bands 1b, 2b, 3b)}	\\
\hline
$\leq 200$ K	&	1	&	(1.52, 1.84, 1.72) & (0.01, 0.08, 0.08)	\\
			&	2	&	(4.84, 5.16, 5.62) & (0.03, 0.11, 0.07)	\\
\hline
$\geq 200$ K	&	1	&	(2.22, 2.08, 2.06) & (0.04, 0.11, 0.28)	\\
			&	2	&	(5.68, 5.16, 4.44) & (0.13, 0.13, 0.24)	\\
\hline
\vspace{0.1cm}\\
\hline \hline
E$\mathu$ range	&	Fit	&	\multicolumn{2}{c}{Gaussian centre [km/s]}	\\
\hline
$\leq 200$ K	&	1	&	(0.02, -0.02, -0.06) & (0.01, 0.03, 0.04)	\\
			&	2	&	(-0.04, 0.03, -0.38) & (0.01, 0.04, 0.04)	\\
\hline
$\geq 200$ K	&	1	&	(0.05, 0.05, -0.22) & (0.02, 0.04, 0.10)	\\
			&	2	&	(0.39, 0.23, 0.04) & (0.05, 0.05, 0.10)	\\
\hline
\end{tabular}
\end{center}
\caption{\changes{Widths and central velocities} of the two Gaussian fit components (``Fit'') to each average line profile in Fig.~\ref{fig:lineprofiles}.}
\label{tab:gaussianfits}
\end{table}

Two-component \changes{blind} Gaussian fits were made for the average profiles in each energy range and band. The \changes{fits} systematically separate into a narrow and broad component\changes{, the widths and central locations of which are given in Table~\ref{tab:gaussianfits}. The superiority of a two-Gaussian fit to that of a single Gaussian, as well as the trends mentioned in the previous paragraph, strongly suggest the presence of at least two physical emission components with different beam filling factors, as discussed in Sect.~\ref{sec:discussion}.}

\subsection{Population diagram and LTE modelling}\label{sec:popdiag}

\begin{figure}[!t]
	\begin{center}
		\includegraphics[clip=,width=1.0\linewidth]{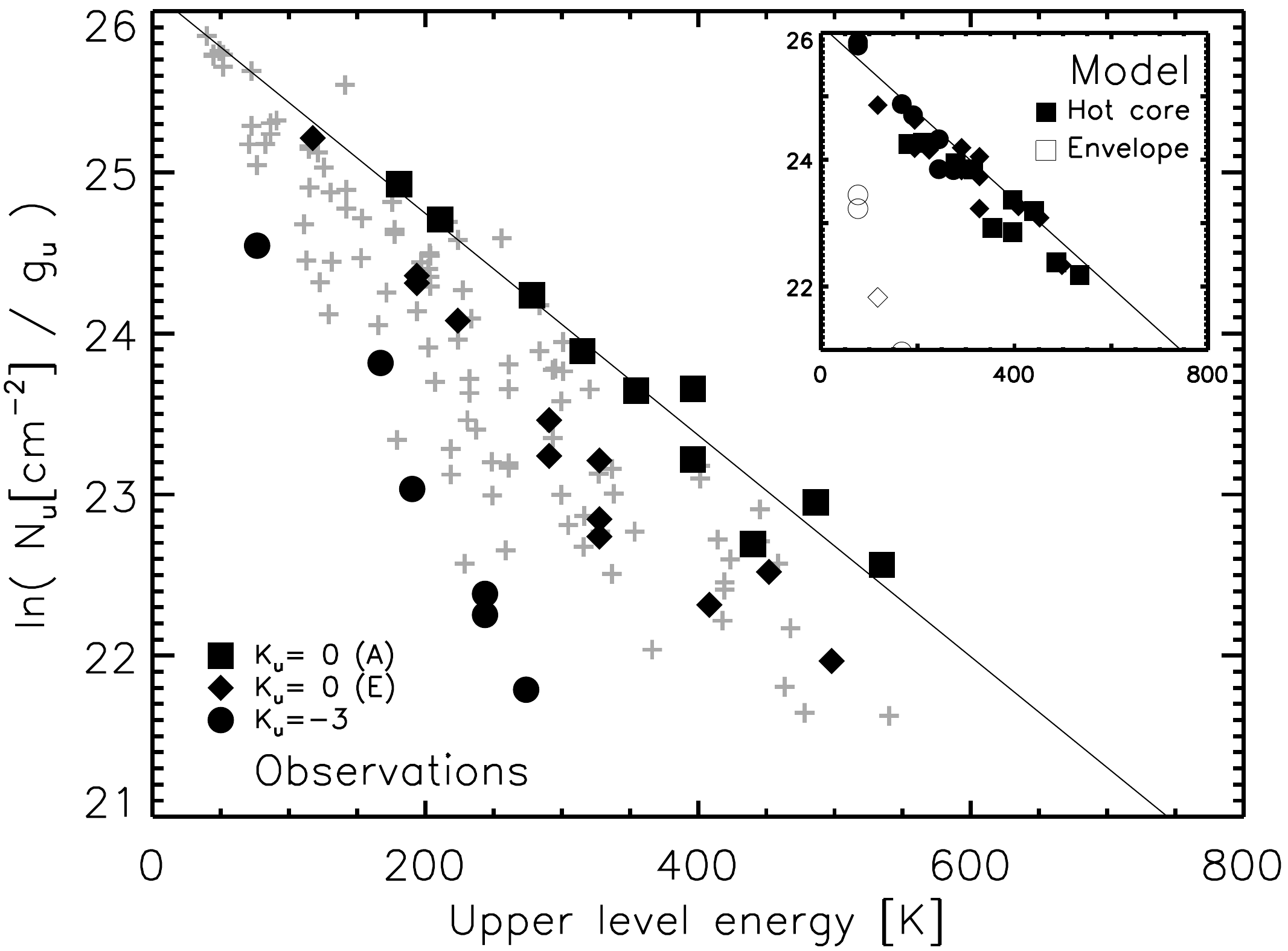}	
		\caption{Observed and modelled (inset) \meth\ population diagrams, plotting column density, divided by statistical weight, versus upper level energy for detected lines from all observed HIFI bands. \changes{Most transitions are given as crosses, but the $K\mathu=0$ ladders for $A$ and $E$ states are indicated by filled squares and diamonds, and $K\mathu=-3$ by filled circles. The inset gives the synthetic population diagram from our LTE model for $K\mathu=0$ and -3, with solid and open symbols representing the hot core and envelope contributions. The linear fit to the observed $K\mathu=0$ $A$ state lines is shown on both plots.}
		}
		\label{fig:popdiag}
	\end{center}
	\end{figure}

The detection of 132 \meth\ lines covering 500K in upper level energy presents an excellent opportunity to perform a population diagram analysis \citep{GoldsmithLanger1999} to study the excitation conditions. In Fig.~\ref{fig:popdiag} we present the population diagram for the observed methanol lines. \changes{Line fluxes were obtained from single-Gaussian fits.}

Part of the \changes{vertical} scatter in Fig.~\ref{fig:popdiag} is due to optical depth, estimated from LTE modelling to contribute $\rm\Delta \ln{(N_{u}/g_{u})}\leq0.37$. \changes{Another factor is beam dilution, which contributes up to} $\rm\Delta \ln{(N_{u}/g_{u})}\approx 2\ln{(37''/25'')}\approx0.8$ for a point source observed at the spectral edges of bands 1b and 3b. However, most of the scatter is due to differences in thermalization conditions for \meth\ transitions with different upper level $K$ quantum numbers. \changes{We assume the K-ladders with the lowest $\rm n_{cr}$, e.g. $K\mathu=0$, approach LTE, as they thermalize below the densities expected for the hot core from the Crim09 model and are seen to lie close to each other, forming the upper envelope of the diagram. The $K\mathu=-3$ ladder, which has a critical density of $\rm n_{cr} \approx 10^{9} cm^{-3}$, forms the lower envelope, well below most detected ladders ($-4 \leq K\mathu \leq 4$).}

\changes{Excitation temperatures and column densities from linear fitting are given in Table~\ref{tab:LTEresults} for the ladder likely closest to LTE, $K\mathu=0$. The fit to the $A$ state is shown in Fig.~\ref{fig:popdiag} and yields \changes{$\rm T_{rot}=(145\pm12)K$} and $\rm N_{A}=(2.2\pm0.5)\cdot10^{14}cm^{-2}$. Fitting the $E$ state gives $\rm T_{rot}=(120\pm8)K$ and $\rm N_{A}=(1.4\pm0.3)\cdot10^{14}cm^{-2}$. A fit to the $K\mathu=-3$ ladder gives $\rm T_{rot}=(66\pm2)K$. This low value is likely mostly due to subthermal excitation, but optical depth and beam dilution may also play a role, as explained earlier.}

\changes{A two-component LTE model was made with the CASSIS software\footnote{CASSIS has been developed by CESR-UPS/CNRS (http://cassis.cesr.fr).}, treating methanol as a single species. The model envelope and hot core have methanol abundances of $2\cdot10^{-9}$ and $10^{-6}$, line widths of \changestwo{3 km/s} and apparent diameters of $20.0''$ and $3.5''$, respectively. Source sizes, H$_{2}$ column densities, and excitation temperatures were taken to be consistent with the Crim09 source model, as well as with the observed $\rm T_{rot}$ values in Table~\ref{tab:LTEresults}, and are discussed in Sect.~\ref{sec:discussion}. Table~\ref{tab:LTEresults} gives the column densities and excitation temperatures we used as input to the model and the simulated observables derived from it. \changestwo{The simulated population diagram, including the \textit{Herschel}-HIFI beam efficiency, is shown as the inset in Fig.~\ref{fig:popdiag}. Optical depth in the modelled lines is typically $0.01\ldots0.80$}. The LTE conditions align transitions from all $K$-ladders, decreasing scatter in the simulated diagram.}

\changes{The model was adjusted to reproduce the $K\mathu=0$ lines, which led to a solution where the hot core dominates the line fluxes, as seen in the inset of Fig.~\ref{fig:popdiag}.} As the  $\rm T_{rot}$ values in Table~\ref{tab:LTEresults} show, a linear fit to the simulated observations yields results different from the model input. This is due to optical depth, \changes{blending}, and beam dilution effects. \changes{The $\rm T_{rot}$ derived from the simulation is sensitive to factor two changes in input line width at the $10\%$ level due to optical depth effects. Owing to the high upper level energies of the detected $K\mathu=0$ transitions, the envelope emission only influences the derived $\rm T_{rot}$ at the 1\% level in the model.} \changes{It is not clear that LTE is a valid assumption even for the $K\mathu=0$ ladder, and thus} a more comprehensive, \changes{non-LTE modelling} effort is underway. Preliminary results suggest the hot core density is higher than in the Crim09 best-fit model.

\begin{table}[!t]
\begin{center}
\begin{tabular}{ r  l  l  l }
\hline \hline
Approach  &  Component  & T [K] (type) & N$\rm_{meth}$[cm$^{-2}$]	\\
\hline
Population & $K\mathu=0 (A)$ & $145\pm$12 (rot) & $(2.2\pm0.5)\cdot 10^{14}$	\\
diagram & $K\mathu=0 (E)$ & $120\pm$8 (rot) & $(1.4\pm0.3)\cdot 10^{14}$	\\
\hline
LTE model & Envelope & $40$ (exc) & $2\cdot 10^{14}$	\\
input & Hot core & $120$ (exc) & $6\cdot 10^{16}$	\\
\hline
LTE simu- & $K\mathu=0 (A)$ & \changes{164} (rot) & $ 1.4\cdot 10^{14}$	\\
lated obs. & $K\mathu=0 (E)$ & \changes{168} (rot) & $ 1.7\cdot 10^{14}$	\\
\hline
\end{tabular}
\end{center}
\caption{From the top: the rotational temperatures and methanol column densities derived from the observed population diagram; the input excitation temperatures and column densities of the LTE model; the rotational temperatures and column densities derived from the modelled population diagram.}
\label{tab:LTEresults}
\end{table}

\section{Discussion}\label{sec:discussion}

The source OMC2-FIR4 is a line-rich protostar with a high degree of chemical and physical complexity. The hundreds of \meth\ transitions that are detectable with HIFI cover a wide range of excitation conditions and offer unprecedented spectral constraints on the structure of protostellar cores.

Previous knowledge \citep[Shim08, Crim09]{Jorgensenetal2006} suggests OMC2-FIR4 is dominated by three components: a large-scale, cool envelope $\sim 10^{4}$AU across; a compact hot core $\sim 10^{2}$AU across; and the outflow from FIR3, in particular a suspected blue-shifted spot resulting from its interaction with the FIR4 envelope. The hot core gives rise to a \changes{dominant part of the emission in all lines in our LTE model, but the changes in line profiles through Fig.~\ref{fig:lineprofiles} imply reality is more complex. The envelope may contribute significantly to the lowest excitation lines, and shocks from regions such as where the FIR3 outflow strikes FIR4 (Shim08) are likely to be important.}

\changes{The effects of the hot core radius and H$_{2}$ column density on the line fluxes are degenerate if the lines are optically thin, as they are in our LTE model. Thus, an upper limit on the former and lower limit on the latter will result in an upper limit on the methanol abundance. Attributing the luminosity of FIR4, $10^{3}$\ \lsol , to the central protostar and assuming grain mantle evaporation at 100 K, one finds an upper limit of $\rm R_{core}\approx760$ AU for $\rm 0.1\mu m$ olivine grains. More realistic radiative transfer would decrease this value. At 440 pc, 760 AU extends $\sim 1.75''$ on the sky, which is adopted as the core radius in the model.}

To reproduce the observed $K\mathu=0$ line fluxes with LTE at an assumed $\rm T_{exc} = 120K$, we need a column density of $\rm N(CH_{3}OH) = 6\cdot10^{16} \textrm{cm}^{-2}$ in the model hot core. Integrating the Crim09 H$_{2}$ density distribution from 100 to \changestwo{R$\rm_{Crim}=$440 AU, one obtains} $\rm N(H_{2}) \approx 6\cdot10^{22} \textrm{cm}^{-2}$. The source model was based on low-resolution maps, and the core is poorly constrained and excludes the inner 100 AU. We thus take the H$_{2}$ column density as a lower limit and, adopting it in the LTE model, conclude that the obtained \meth\ abundance $\rm X_{core}=10^{-6}$ is a conservative upper limit. The highest abundance seen in low-mass hot cores is $2\cdot10^{-6}$ \citep{Jorgensenetal2005}. If the hot core does not dominate the line emission in reality, smaller hot core sizes and methanol column densities, as well as higher rotational temperatures, may be consistent with the data. Presently, the abundance of methanol in the envelope is very poorly constrained. We set it to $2\cdot10^{-9}$ in the LTE model, consistent with the factor $10^{2\ldots3}$ abundance jump seen in hot cores \citep{vanderTaketal2000, Maretetal2005}.

\changes{The hot core density is broadly constrained to be between $10^{6}$ and $\rm 10^{8}\ cm^{-3}$ by this early analysis. Lower and higher values would be difficult to reconcile with the large-scale density profile, furthermore the population diagram suggests the $K$ ladders with the lowest $\rm n_{cr}$ may approach LTE while those with the highest $\rm n_{cr}$ clearly deviate from it. Typical kinetic temperatures well in the $\rm\geq 100$ K hot core regime are implied by the obtained rotational temperatures of $\rm T_{rot} \approx (120\ldots145)$ K. Shock contributions from the FIR3-FIR4 interaction spot as well as the hot core itself need to be considered before drawing further conclusions about the temperature structure.}

An intriguing feature is the appearance of an asymmetry in the \changes{$\rm E_{u}\geq 200$ K lines, as seen in Fig.~\ref{fig:lineprofiles}}. Because these lines should be dominated by the hot core emission, the profiles observed could represent the blue asymmetric profile expected for infalling gas \citep[e.g.][]{Walkeretal1994}, implying we \changes{may be} seeing collapse in the central regions. \changes{To clarify whether infall, shocks, or other phenomena are responsible for the asymmetry, interferometry} as well as further analysis of the \changes{\textit{Herschel}-HIFI data, in particular the $^{13}$\meth\ lines, will be employed.}

\changes{We note the difference in column density of the $K\mathu=0$ transitions of the $A$ and $E$ states in Table~\ref{tab:LTEresults}. Whether it is truly significant will be explored in a future paper. Also, early results from an analysis of formaldehyde in OMC2-FIR4 will be presented in a companion paper (Crimier et al., in prep.).}

\begin{acknowledgements}
The authors are grateful to the referee, Dr. Tim van Kempen, for constructive comments leading to a significant improvement of the paper, and to Rens Waters for helpful discussions. 
HIFI has been designed and built by a consortium of institutes and university departments from across 
Europe, Canada and the United States under the leadership of SRON Netherlands Institute for Space 
Research, Groningen, The Netherlands and with major contributions from Germany, France and the US. 
Consortium members are: Canada: CSA, U.Waterloo; France: CESR, LAB, LERMA,  IRAM; Germany: 
KOSMA, MPIfR, MPS; Ireland, NUI Maynooth; Italy: ASI, IFSI-INAF, Osservatorio Astrofisico di Arcetri- 
INAF; Netherlands: SRON, TUD; Poland: CAMK, CBK; Spain: Observatorio Astron—mico Nacional (IGN), 
Centro de Astrobiolog'a (CSIC-INTA). Sweden:  Chalmers University of Technology - MC2, RSS \& GARD; 
Onsala Space Observatory; Swedish National Space Board, Stockholm University - Stockholm Observatory; 
Switzerland: ETH Zurich, FHNW; USA: Caltech, JPL, NHSC, and we are deeply grateful to everyone involved in the designing, building, and exploitation of this fantastic instrument.
HCSS, HSpot, and HIPE are joint developments by the \textit{Herschel} Science Ground 
Segment Consortium, consisting of ESA, the NASA \textit{Herschel} Science Center, and the HIFI, PACS, and 
SPIRE consortia. 
M.Kama gratefully acknowledges support from the Netherlands Organisation for Scientific Research (NWO) grant number 021.002.081 and the Leids Kerkhoven-Bosscha Fonds, and thanks SRON Groningen for hosting the HIFI ICC volunteers. 
\end{acknowledgements}

\bibliographystyle{aa}
\bibliography{omc2fir4_methanol}

\end{document}